\documentclass[twocolumn,10pt]{amsart}
\usepackage[hidelinks,bookmarks=false]{hyperref}
\usepackage{graphicx}
\usepackage{xcolor}
\usepackage{natbib}
\usepackage{blindtext}

\usepackage{caption}

\newcommand{\plotdir}{./plots/}

\usepackage{geometry}
\usepackage{algorithm}
\usepackage{algorithmicx}
\usepackage{bbm}

\newcommand{\vp}{\varphi}

\newcommand{\rvp}{r(\varphi)}
\newcommand{\rvph}{\hat{r}(\varphi)}
\newcommand{\wkvp}{w_k(\vp)}
\newcommand{\wkvph}{\hat{w}_k(\vp)}
\newcommand{\wvph}{\hat{w}(\vp)}

\newcommand{\figw}{.5\textwidth}

\begin{document}

\twocolumn[\title{Support Vector Spectrum Approximations: Efficient Calculations of Light Scattering Lineshapes}

\author{Gregory J. Hunt$^{1,*}$, 
  Cody R. Ground$^{2}$, 
  and Robin L. Hunt$^{2}$\\
  $^{1}$William \& Mary, PO Box 8763 Williamsburg, VA 23187\\
$^{2}$Hypersonic Airbreathing Propulsion Branch, NASA Langley Research Center, 12 Langley Blvd. Hampton, VA 23681.\\
$^{*}$ghunt@wm.edu
}
\maketitle


\begin{abstract}
Measuring scattered light is central to many laser-based gas diagnostic techniques, e.g., coherent anti-Stokes Raman spectroscopy (CARS) and filtered Rayleigh scattering (FRS). To produce quantitative measurements with such techniques, a computational model of the scattered spectral lineshape is necessary. While accurate, these models are often quite computationally demanding and thus cannot be used in situations where computational speed matters. To overcome this, approximations of these spectral lineshape models can be used instead. In this paper, we develop a method called support vector spectrum approximation (SVSA). This method uses machine learning to create efficient and accurate approximations of any existing spectral lineshape model. The SVSA framework improves upon existing methods by allowing efficient approximations of spectral lineshapes to be calculated in arbitrary flow regimes. We demonstrate the efficacy of SVSA in approximating coherent and spontaneous Rayleigh-Brillioun spectra. We also show that SVSA reduces the computational cost of a simulated filtered Rayleigh scattering experiment by a factor of 300.
\end{abstract}]

\section{Introduction}

Measuring scattered light is central to many laser-based gas diagnostic techniques. For example, coherent anti-Stokes Raman spectroscopy (CARS) and filtered Rayleigh scattering (FRS) are techniques using Raman and Rayleigh\allowbreak-Brillouin  (RB) scattering, respectively, to quantify properties of a gas such as pressure, velocity, temperature, and number density \citep{Boguszko2005, Miles2001, Grant2000}. Such laser-based gas diagnostic techniques are useful in a diverse set of fields. In atmospheric science, measurements of RB scattered light are used for remote sensing of clouds, aerosols, winds, ice crystals, etc. \citep{Witschas2014, Binietoglou2016}. In the physics and engineering fields, Raman and RB scattering are used to nonintrusively investigate flames and other complex flows \citep{Doll2017,Miles2001, Ehn2017}.

To produce quantitative measurements using these techniques, one often needs to know the spectral lineshape of the scattered light. This lineshape conveys important information about the gas' properties. Consequently, accurate computational models linking gas properties to the spectral lineshape have been developed in the literature. For example, CARSFT can be used to calculate coherent anti-Stokes Raman spectra and Tenti's S6 model or Pan's S7 model allow calculation of RB spectra \citep{Palmer1989,Tenti1974, Pan2002}. While typically quite accurate, such lineshape models are often computationally demanding. Thus, they are not ideal for situations where computational speed matters, for example, in optimization studies or real-time diagnostic techniques \citep{Yeaton2012,Binietoglou2016}.

For RB scattering, this problem has motivated methods to efficiently approximate the RB spectral lineshape. In particular, approximating Tenti's S6 model has been previously investigated \citep{Witschas2014, Ma2012, Binietoglou2016}. Such methods have had good success in the regime on which they are designed. However, there does not presently exist a general approach for rapidly approximating general spectral lineshapes for any given scattering process over an arbitrary regime. Existing approximation frameworks are specific to particular types of scattering, like Rayleigh\allowbreak-Brillouin, and cannot be used to produce approximations of, for example, Raman spectra. Similarly, methods may only be valid for a small range of parameters or they may simplify the lineshape model by treating parameters as constant even though they are not. Consequently, while these methods may be accurate for, say, atmospheric regimes, they may not transfer well to highly dynamic conditions like those in supersonic flows.

This paper proposes a method called support vector spectrum approximation (SVSA), that creates efficient and accurate approximations of any spectral lineshape given a preexisting computational model. The method applies machine learning to produce a general lineshape approximator. The framework is applicable to preexisting lineshape models of any scattering type in an arbitrary flow regime. We show that SVSA is both fast and accurate by demonstrating its efficacy approximating coherent and spontaneous RB spectra. Furthermore, in application on a real problem, we demonstrate how SVSA might be used in an FRS study of hypersonic vehicle engines to make the computations feasible. 

\section{Methods}\label{sec:methods}

\begin{figure}
  \centering
  \includegraphics[width=\figw]{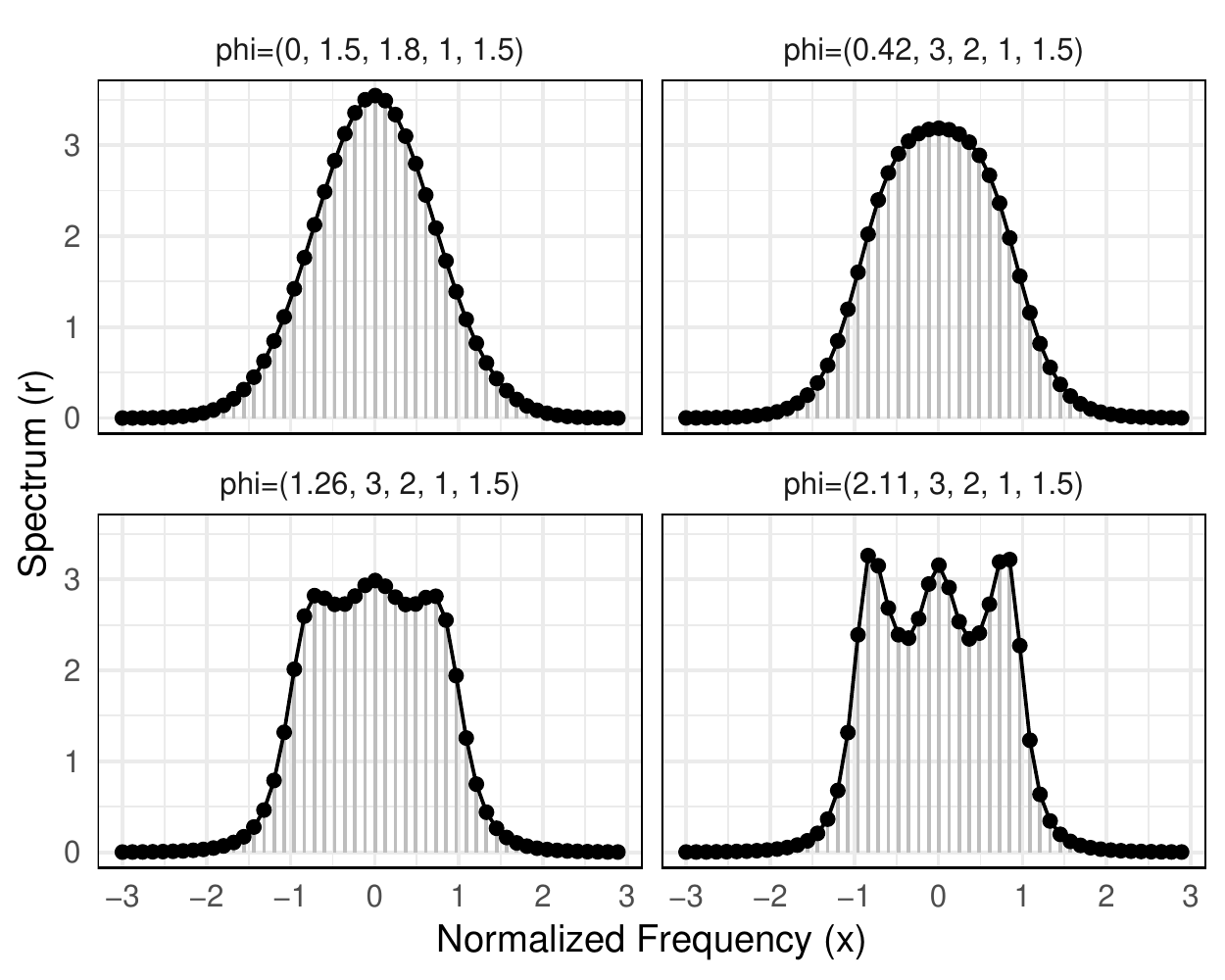}
  \caption{Examples of spontaneous Rayleigh-Brillouin spectral lineshapes for different sets of input parameters calculated using Tenti's S6 model. Here, the order of $\vp$ is y, internal relaxation number, Euken factor, internal specific heat, and translational specific heat.}\label{fig:ex_lines}
\end{figure}

Given any preexisting lineshape model, we generically denote the set of input parameters for the model as $\varphi\in\mathbb{R}^P$ where $P$ denotes the number of parameters. For example, $P=5$ for the S6 and S7 models of RB scattering, which have as parameters y, Euken factor, internal relaxation number, internal specific heat, and translational specific heat. Given $M$ values of the (normalized) frequency $x_1,\ldots,x_M$, a lineshape model $r$ takes $\varphi$ and produces an estimate $r(\varphi) \in \mathbb{R}^{M}$ of the spectrum at $x_1,\ldots,x_M$. Figure~\ref{fig:ex_lines} illustrates several examples of spontaneous Rayleigh-Brillouin scattering spectra estimated by Tenti's S6 model.

 The goal of SVSA is to create fast approximations of $r(\varphi)$. SVSA does this by leveraging machine learning. Specifically, the method combines support vector regression (SVR) and the singular value decomposition (SVD) to build a flexible and general approximator.

\subsection{The SVSA Model}

At the heart of SVSA is an assumption that most of the variability among spectral lineshapes can be explained by a small number of dominant lineshape modes denoted $f_1,\ldots,f_K$. That is, for any $\vp$, we assume that 
\begin{equation}\label{eqn:sva_approx}
\rvp \approx \sum_{k=1}^{K}w_k(\vp)f_k
\end{equation}
so that a lineshape is approximately the linear combination of some dominant global modes $f_1,\ldots,f_K$ with parameter-specific weights $w_k(\vp)$. Using this approximation, SVSA estimates $\rvp$ by estimating the $\{f_k\}$ and the $\{w_k(\vp)\}$ and combining them according to Equation~\eqref{eqn:sva_approx}.

Given training data generated by a preexisting computational lineshape model, SVSA does this estimation through a two-step approach:
\begin{enumerate}
    \item Decompose the training data with the SVD to estimate $K$ functional modes $\hat{f}_1,\ldots,\hat{f}_K\in\mathbb{R}^M$. 
\item Learn the relationship between $\vp$ and $\wkvp$ using SVR. This produces, for each $k$, a model $\hat{w}_k$ where $\wkvph\approx \wkvp$. 
\end{enumerate}

Combining the estimates from these two steps, SVSA predicts $\rvp$ as
\begin{equation}\label{eqn:rhat_interp}
\rvph = \sum_{k=1}^{K}\wkvph \hat{f}_k.
\end{equation}
An important feature of SVSA is that its calculation is fast. We show in the next section that $\rvph$ can be succinctly written as 
\begin{equation}\label{eqn:rhat_expand}
\rvph = \beta_0 + \beta z(\vp) 
\end{equation}
with
\begin{equation}
  \label{eqn:z}
  \begin{aligned}
    z(\vp) &= \exp\left(-||\Sigma-\vp||^2\right)\\
    \end{aligned}
\end{equation}
where $\beta_0\in\mathbb{R}^{M}$ and $\beta\in\mathbb{R}^{M\times S}$ are the model coefficients from the SVR models expressed in the $\{\hat{f}_k\}$ basis and $\Sigma\in\mathbb{R}^{S\times P}$ is a matrix with rows $\{\Sigma_s\}_{s=1}^{S}$ that are ``support vectors'' $\Sigma_s\in\mathbb{R}^P$ chosen by SVR. The notation ``$\Sigma-\vp$'' in Equation~\eqref{eqn:z} denotes the $S \times 1$ vector where
\[
\left(\Sigma-\vp\right)_s = \Sigma_s-\vp.
\]
Notice that we can expand $-||\Sigma-\vp||^2$ as
\[
-(\Sigma-\vp)'(\Sigma-\vp) = 2\Sigma\vp - \text{diag}^{-1}\left(\Sigma'\Sigma\right) - \vp'\vp
\]
where the $\text{diag}^{-1}$ operator forms the vector of diagonal elements of a matrix. Thus, calculating $\rvph$ boils down to matrix multiplication and exponentiation. This means SVSA can be calculated very efficiently using standard scientific computing software. 

\subsection{Fitting SVSA}

\begin{figure}\centering
\includegraphics[width=\figw]{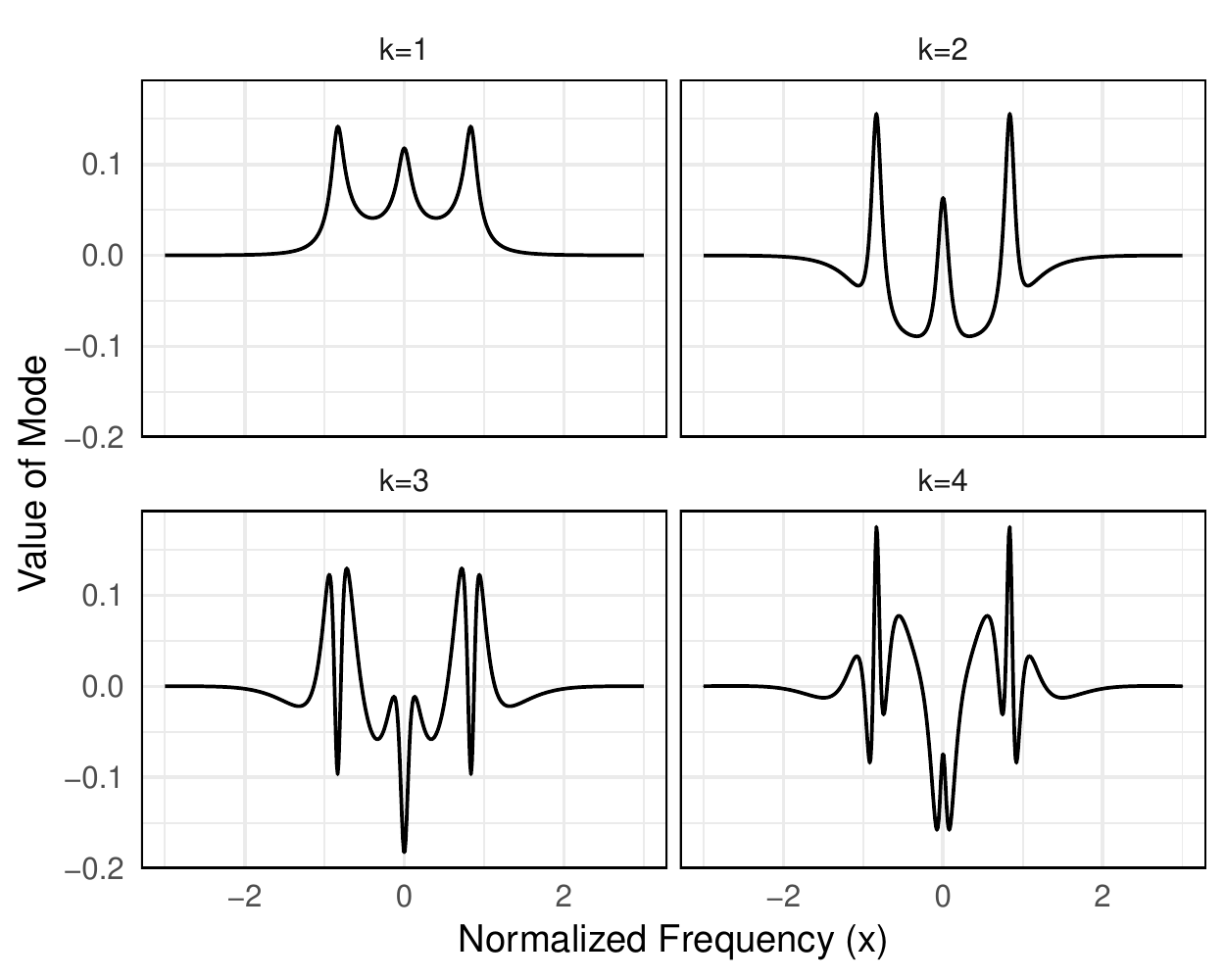}\caption{Dominant four lineshape modes for example training data.}\label{fig:uk}
\end{figure}

The SVSA method learns its model parameters $\beta_0,\beta$ and $\Sigma$ using a set of $N$ training examples. Let $\varphi_1,\ldots,\varphi_N\in\mathbb{R}^P$ be a collection of training parameters that cover the range of the input space over which we want to learn the approximator and let $r(\varphi_1),\ldots,r(\varphi_N)\in\mathbb{R}^{M}$ be the corresponding training lineshapes calculated by our existing lineshape model. The first task when fitting SVSA is to learn the dominant functional modes $f_1,\ldots,f_K$. We do this with the SVD. Let $R\in\mathbb{R}^{M \times N}$ be the lineshape matrix containing the $N$ training spectra so that $R_{mn}$ is the value of  $r(\vp_n)$ at $x_m$. If $R$ has an SVD of $R=UDV'$, then define $W=DV'$ so that $R=UW$. This factorization allows us to decompose the $n^{th}$ column of $R$ as 
\begin{equation}\label{eqn:decomp}
r(\vp_n) = \sum_{k=1}^{M} W_{kn}U_{k}
\end{equation}
where $U_k$ is the $k^{th}$ column of $U$. These columns of $U$ (the noncentered principal components of the training lineshapes) capture the dominant functional modes of the lineshapes in our training data. Figure~\ref{fig:uk} displays the top four functional modes of some example training data. The first mode approximately captures the average lineshape of the training data and subsequent modes capture typical deviations from this average.

Since the $\{U_k\}$ are ordered decreasing according to how much variation they explain, if we truncate the linear combination in Equation \eqref{eqn:decomp} after $K<M$ terms, then $\sum_{k=1}^{K} W_{kn}U_k$ is the best approximation of $r(\vp_n)$ using only $K$ modes. We thus let $\hat{f}_k=U_k$ for $k=1,\ldots,K$ and choose $K$ so that these $\{\hat{f}_k\}$ capture most of the variation present in the training lineshapes. We can often achieve a very good approximation using $K \ll M$. We explore the accuracy of this approximation in Section~\ref{sec:eval}.

The second step to training SVSA is to learn the relationship between $\vp$ and $w_k(\vp)$. To do this, we need training examples of the $\vp$ and corresponding weights $w_k(\vp)$ from which to learn. We can reuse, the training data used to estimate the $\{f_k\}$ by noting that for each training parameter $\vp_n$ the correpsonding weight $w_k(\vp_n)$ is precisely $W_{kn}$. To see this, consider the similar form of Equations~\eqref{eqn:sva_approx}, \eqref{eqn:rhat_interp} and \eqref{eqn:decomp}. The $\{W_{kn}\}_{k=1}^K$ in Equation~\ref{eqn:decomp} are essentially the weights $\{w_k(\vp_n)\}_{k=1}^K$ used to approximately reconstruct $r(\vp_n)$ from $f_1,\ldots,f_K$ (Equation~\eqref{eqn:sva_approx}). Thus, to learn the $w_k$, we use the training parameters $\vp_n$ and corresponding ``training weights'' $W_{kn}=w_k(\vp_n)$ to train $K$ different SVR models. That is, for each $k=1,\ldots,K$, we find a fit to predict $W_{kn}$ from $\vp_n$. We call this fit $\hat{w}_k$ so that $\hat{w}_k(\vp_n)\approx w_k(\vp_n)=W_{kn} $.

The SVR approach used for this fit finds a flexible and regression-like function to predict $w_k(\vp)$ from $\vp$. Fitting these models amounts to using the training data to identify an important subset of training points $\Sigma_{1},\ldots,\Sigma_{S}\in\mathbb{R}^P$, called support vectors, and an associated set of coefficients $\alpha_{k0},\ldots,\alpha_{kS}\in\mathbb{R}$. The SVR models use these parameters to predict $w_k(\vp)$ from $\vp$ as a linear combination of the $\{\alpha\}$ weighted by the similarity between $\vp$ and $\Sigma_{s}$. Specifically,
\begin{equation}\label{eqn:svr}
  \hat{w}_k(\vp)=\alpha_{k0}+\sum_{s=1}^{S}\alpha_{ks}H(\Sigma_{s},\vp)
\end{equation}
where $H$ is a kernel function measuring the similarity between $\Sigma_s$ and $\vp$. Essentially, the SVR models use the support vectors as a set of important landmarks in the parameter space of all $\vp$ and estimates $w_k(\vp)\approx w_k(\Sigma_s)$ to the extent that $\vp\approx \Sigma_s$.

Using SVR-based estimators over more traditional approaches like multiple regression has substantial advantages. First, SVSA uses $\nu$-SVR, a version of SVR parameterized by $\nu \in [0,1]$. This type of SVR finds an optimal fit for the data that uses (approximately) $\nu$ percentage of the training points as support vectors. Thus, $\nu$ directly controls the model complexity and allows for an intuitive trade-off between accuracy and speed. A low value of $\nu$ forces the fit to only retain the few most important support vectors. In this case, predictions will be less computationally demanding by sacrificing accuracy. Conversely, a high value of $\nu$ means the model can keep track of many support vectors and build a nuanced and highly accurate model. However, this prediction accuracy comes at the cost of computational speed.

A second advantage to SVR is our use of a radial basis function (RBF) kernel $H$ in Equation~\eqref{eqn:svr} defined as $H(x,y) = \exp(-||x-y||^2)$ for $x,y\in\mathbb{R}^P$. This allows SVR to fit highly nonlinear relationships between $\vp$ and $w_k(\vp)$. While multiple regression would find a fit that is linear in $\vp$, RBF kernelized SVR amounts to fitting a regression-like estimator in an expanded infinite dimensional space of all possible polynomials of $\vp$. Thus, SVSA's approach allows an increased amount of flexibility in the model fit. It automatically considers all possible polynomial relationships between $\vp$ and $w_k(\vp)$ and finds the best fit among those while avoiding overfitting.

Finally, once fit, SVSA is very fast to compute. If $\hat{w}(\vp)=(\hat{w}_1(\vp),\ldots,\hat{w}_k(\vp))$ then we can succinctly write $\hat{w}(\vp) = \alpha_0 + \alpha z(\vp)$ with $z$ as in Equation~\eqref{eqn:z}, where $\alpha_0\in\mathbb{R}^K$ and $\alpha\in\mathbb{R}^{K\times S}$ contain the coefficient estimates across all $K$ of our SVR models. The parameter $\Sigma$ in $z$ (see Equation~\ref{eqn:rhat_expand}) is the combined matrix of all support vectors showing up in any of our models. Here $\Sigma\in\mathbb{R}^{S\times P}$ where $S$ is the total number of unique support vectors across the models. Letting $\hat{F}=\left[\hat{f}_1|\cdots|\hat{f}_K\right]\in\mathbb{R}^{M\times K}$, we have
\[
\rvph = \hat{F}\wvph = \hat{F}\left(\alpha_0 + \alpha z(\vp)\right) = \beta_0 +\beta z(\vp)
\]  where we define $\beta_0 = \hat{F}\alpha_0$ and $\beta = \hat{F}\alpha$.
Thus, the major components of calculating $\rvph$ are (1) calculating $Z(\vp)$, which mainly involves the exponential and the product $\Sigma\vp$, and (2) calculating a product between $\beta$ and $Z(\vp)$. More generally, if we have a matrix $\Phi=\left[\vp_1 | \vp_2 | \cdots | \vp_L \right]\in\mathbb{R}^{P\times L}$ consisting of parameter values $\vp_1,\ldots,\vp_L\in\mathbb{R}^P$ for $L$ lineshapes, then we can efficiently approximate all of the lines as $\hat{R}(\Phi)\in\mathbb{R}^{M \times L}$ letting
\[
\hat{R}(\Phi) = \beta_0 + \beta Z(\Phi)
\]
with
\[
\begin{aligned}
  Z(\Phi)&=\exp\left(-||\Sigma-\Phi||^2\right).
\end{aligned}
\]
Here, $\Sigma-\Phi$ denotes an $S \times L$ matrix where
\[
(\Sigma-\Phi)_{s\ell} = \Sigma_s - \vp_\ell,
\]
the the norm $||\cdot||$ is applied columnwise and the exponential elementwise. As $Z$ is a Gaussian kernel, it must be bounded for all $\Phi$ and cannot be ill-conditioned.  Again, notice that we can write the exponent as
\[-||\Sigma-\Phi||^2 = 2\Sigma\Phi - \text{diag}^{-1}(\Sigma'\Sigma) - \text{diag}^{-1}(\Phi\Phi')\] where the first subtraction is rowwise and the second is columnwise.  

All together, after estimating $\beta_0, \beta$ and $\Sigma$ using the training data, SVSA boils down to the following three steps codified in Algorithm~\ref{alg:svsa}:

\begin{algorithm}[H]\caption{Support Vector Spectrum Approximation}\label{alg:svsa}
  \begin{algorithmic}
    \State {\bf Step 1.} Calculate \[
    \begin{aligned}
      E&=-||\Sigma-\Phi||^2\\
      &= 2\Sigma\Phi - \text{diag}^{-1}(\Sigma'\Sigma) - \text{diag}^{-1}(\Phi\Phi').
    \end{aligned}
    \]
\State {\bf Step 2.} Exponentiate elements of $E$ to get $Z(\Phi)$.
\State {\bf Step 3.} Calculate $\hat{R}(\Phi) = \beta_0 + \beta Z(\Phi)$. 
  \end{algorithmic}
\end{algorithm}

The computationally expensive parts of each step are (1) calculating $E$ in $\mathcal{O}\left(SPL\right)$ operations, (2) exponentiating the elements of $E$ using $SL$ exponentiations, and (3) calculating $\beta Z(\Phi)$ with $\mathcal{O}\left(MSL\right)$ operations. As each of these steps are fast to compute using modern scientific computing packages, SVSA allows us to very efficiently compute a highly flexible, nonlinear approximation of $\rvp$ with basically just two matrix products and exponentiation. Code for fitting and applying our method can be found on github at \url{gjhunt.github.io/svsa}. Currently, the fitting and lineshape approximation algorithms are implemented in {\tt Python}. However, once the coefficients have been estimated in {\tt Python}, they may also be exported and used to approximate the lineshape using routines we have written in {\tt Matlab} and {\tt R}. 

\section{Evaluation on Data}\label{sec:eval}

To assess the accuracy and speed of SVSA, we generate example lineshape data on which to evaluate our method. We generate spontaneous Rayleigh-Brillouin scattering (RBS) spectral lineshapes according to Tenti's S6 model \citep{Tenti1974}. Tenti's model estimates RBS using the parameters: y, Euken factor, internal relaxation number, internal specific heat and translational specific heat. Our primary interest in SVSA is application to a supersonic wind tunnel where a shock wave system is generated to simulate the flow conditions in a high speed air-breathing engine. Thus, we generated our training lineshapes using a range of parameters typically seen in these wind tunnel studies. We set the internal specific heat ({\tt c\_int}) to $1$, the translational specific heat ({\tt c\_tr}) to $3/2$, and vary $P=3$ parameters: the Euken factor ({\tt Eukenf}) over a grid of 5 points from $1.8$ to $2$, the internal relaxation number ({\tt rlx\_int}) over a grid of 5 points from $1.5$ to $3$, and the $y$ factor ({\tt y}) over a grid of 20 points from 0 to 8. In total, this generates a set of $500$ training lineshapes. Each lineshape is calculated on a grid of $500$ values of the nondimensionalized frequency ({\tt x}) ranging from $-3$ to $3$.

\begin{figure}
  \centering
  \includegraphics[width=\figw]{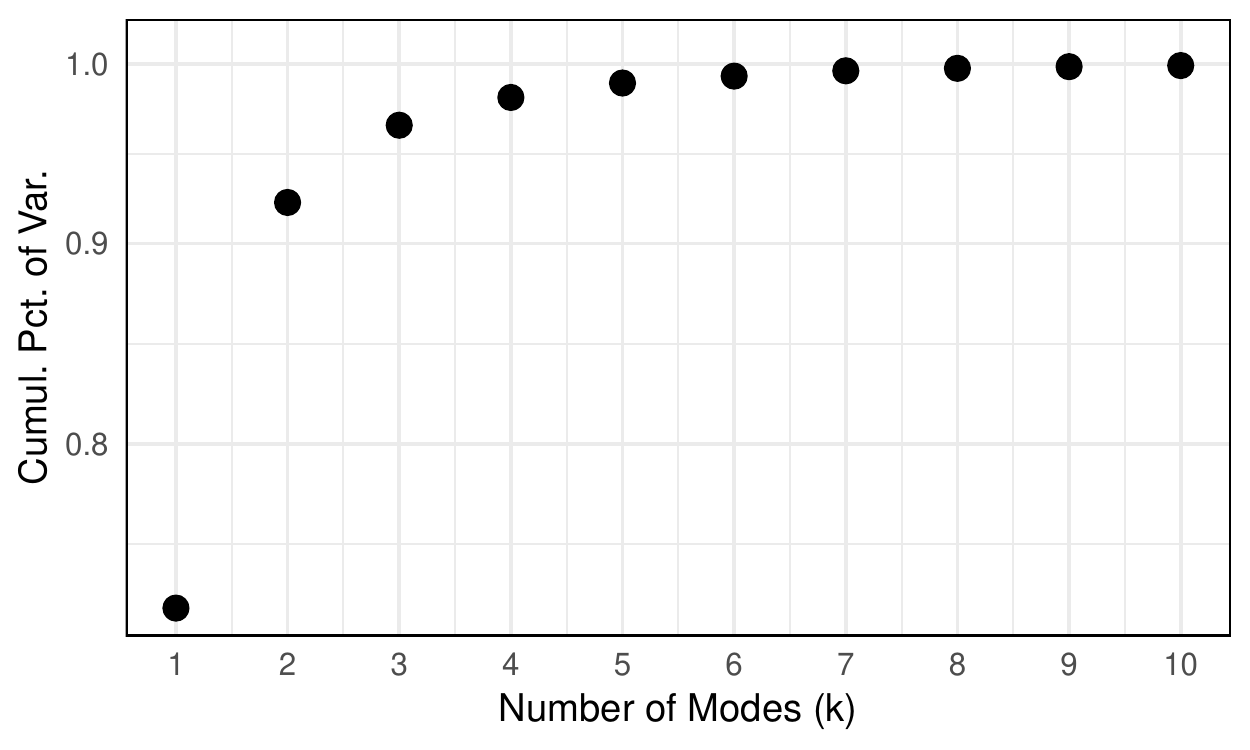}
  \caption{Cumulative percentage of variance captured by first several dominant modes of the training data.}\label{fig:scree}
\end{figure}

In Figure~\ref{fig:scree}, we plot the cumulative percentage of variance captured in the training data by the first several estimated modes $\hat{f}_k$. 
We can see from this figure that almost all of the variability among spectral lineshapes can be captured using the first several dominant modes. By approximating the spectral lineshapes using $K\approx 6$ modes, we are able to recover $>99\%$ of the important differences among lineshapes. The dimensionality reduction achieved by these modes makes SVSA efficient to compute since we need only train $6$ SVR models.

\begin{figure}
  \centering
  \includegraphics[width=\figw]{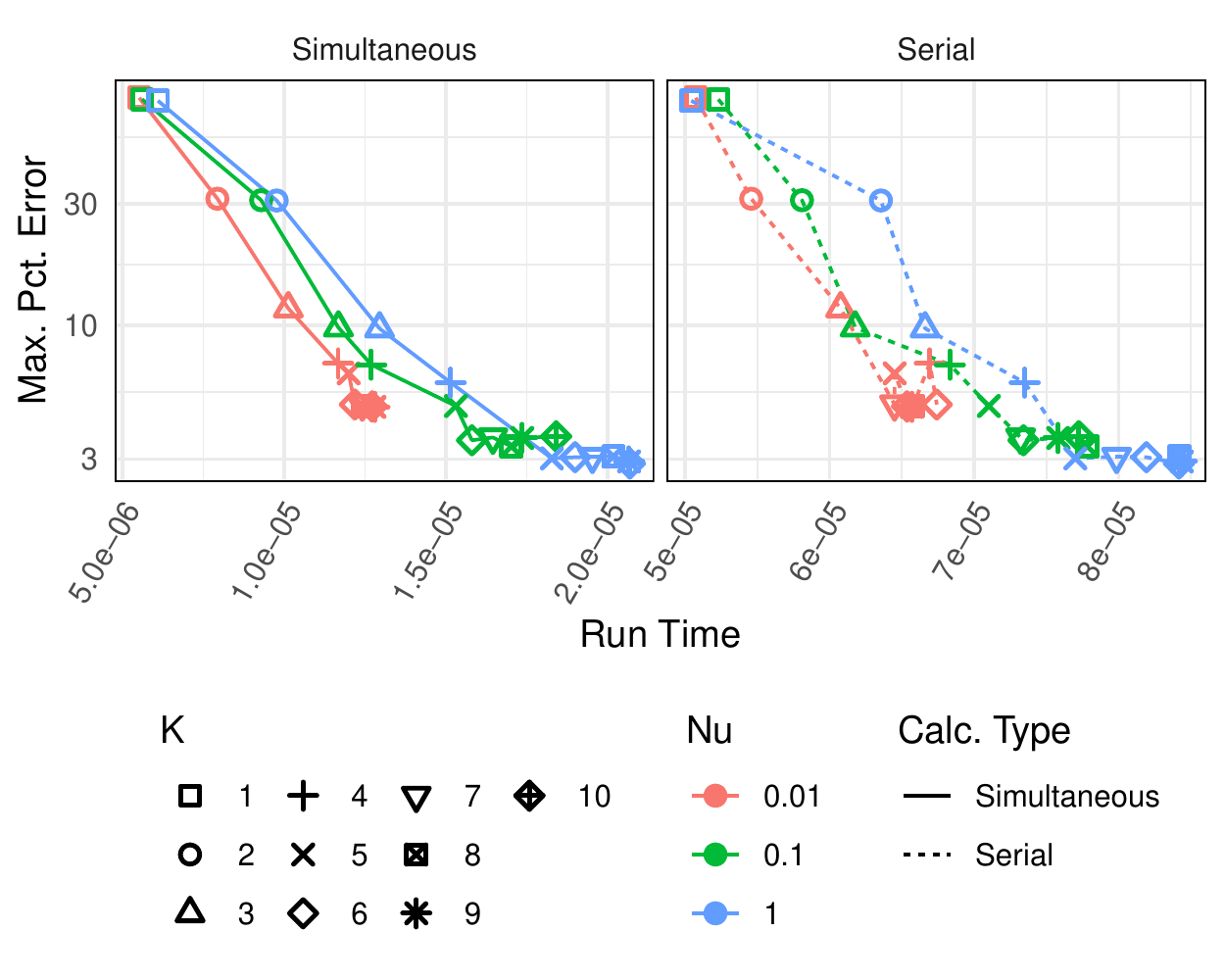}
  \caption{Percent error by average computation time for 32,000 test lineshapes calculated simultaneously and serially.}
  \label{fig:err_nuk}
 \end{figure}

To further explore the relationship between model complexity and accuracy, we generate a collection of test lineshapes covering the same parameter ranges as the training data but over a finer grid. For the test data, we generate 32,000 example lineshapes. In Figure~\ref{fig:err_nuk}, we plot the maximum error of SVSA (across all 32,000 test lineshapes) against the average time needed to calculate a lineshape using the method. We do this for three values of $\nu$, ten values of $K$, and two ways of calculating lineshapes.

 The $y$-axis in Figure~\ref{fig:err_nuk} measures the maximum difference between the SVSA approximated lineshape and the true lineshape as a percentage of the maximum height of the true line. The $x$-axis in Figure~\ref{fig:err_nuk} measures the time it takes SVSA to make the predictions. We display two sets of curves for two ways of calculating the 32,000 lineshapes. First, the lineshapes are calculated by predicting all 32,000 lineshapes simultaneously. This is done by feeding SVSA a matrix $\Phi \in\mathbb{R}^{3 \times 32000}$ of all parameters as per Algorithm~\ref{alg:svsa}. We divide the total time it takes to make all predictions by 32,000. Secondly, we calculate the test lineshapes in a serial fashion feeding one parameter vector $\vp\in\mathbb{R}^{3\times 1}$ to SVSA at a time. We then loop over all 32,000 parameters and report the average computation time. These two ways of calculating the lineshapes cover two common use-cases for SVSA. The first is the batch setting where all of the parameters are known in advance of the calculation. For example, this occurs when analyzing data collected from an experiment or a computational fluid dynamics (CFD) simulation. The second situation covers doing calculations in a streaming context where new data arrives over time. In this case, we do not know the parameter values in advance and thus, can't calculate them in bulk. We must approximate each lineshape separately as the data arrives. This is what our loop over the 32,000 test cases demonstrates. An example of this situation is real-time laser diagnostics. For both approaches, the displayed time in Figure~\ref{fig:err_nuk} is the average number of (wall-time) seconds it takes our Python implementation to make a prediction on a 4.2 GHz processor.

Figure~\ref{fig:err_nuk} shows a large difference between estimating the lineshapes simultaneously and estimating them serially. It is almost an order of magnitude faster to simultaneously calculate the lineshapes. This happens because approximating the lineshapes simultaneously allows the entire calculation to be handled extremely efficiently by low-level linear algebra packages. Alternatively, the overhead when calculating the lineshapes serially slows down the average approximation speed. Nonetheless, in either case, SVSA is still extremely fast. Tenti's model takes about one-tenth of a second to calculate a single lineshape on our 4GHz processor. Thus, depending on the parameter setttings, SVSA is between one-thousand and one-hundred-thousand times faster.

We also notice in Figure~\ref{fig:err_nuk} that both $K$ and $\nu$ have substantial impacts on both speed and accuracy. Increasing either $K$ or $\nu$ increase the model complexity, and thus decrease error at the cost of speed. For $\nu$, this effect is straight-forward. Increasing $\nu$ increases the number of support vectors. This leads to more nuanced and accurate models that are slower to compute. For $K$, larger values indirectly affect model complexity by increasing the complexity of the SVR models we build. This is because the weight functions $w_k$, corresponding to a mode $f_k$, are increasingly complex as we increase $k$ since higher-order $f_k$ pick up more subtle and complicated modes. Modeling these complicated modes requires more flexible models and hence more support vectors. Thus, increasing $K$ correspondingly increases model complexity which decreases error at the expense of speed. Notice, however, increasing $K$ quickly reaches diminishing returns. Indeed using more than about $6$ modes seems to only increase computation time. This is because the higher order modes explain an ever-smaller proportion of the variability among the lineshapes. For example, from Figure~\ref{fig:scree}, we see that the $7^{th}$ mode accounts for less than $0.003\%$ of each line's shape. Thus, including these higher order terms increases model complexity for relatively little gain in approximation accuracy. For our application, it seems that we do well using $K=6$ and $\nu=0.1$. This yields a maximum error in our test set of about 4\% and an average computation time between about $1\times 10^{-5}$ and $8\times 10^{-5}$ seconds per line.

\begin{figure}\centering
  \includegraphics[width=\figw]{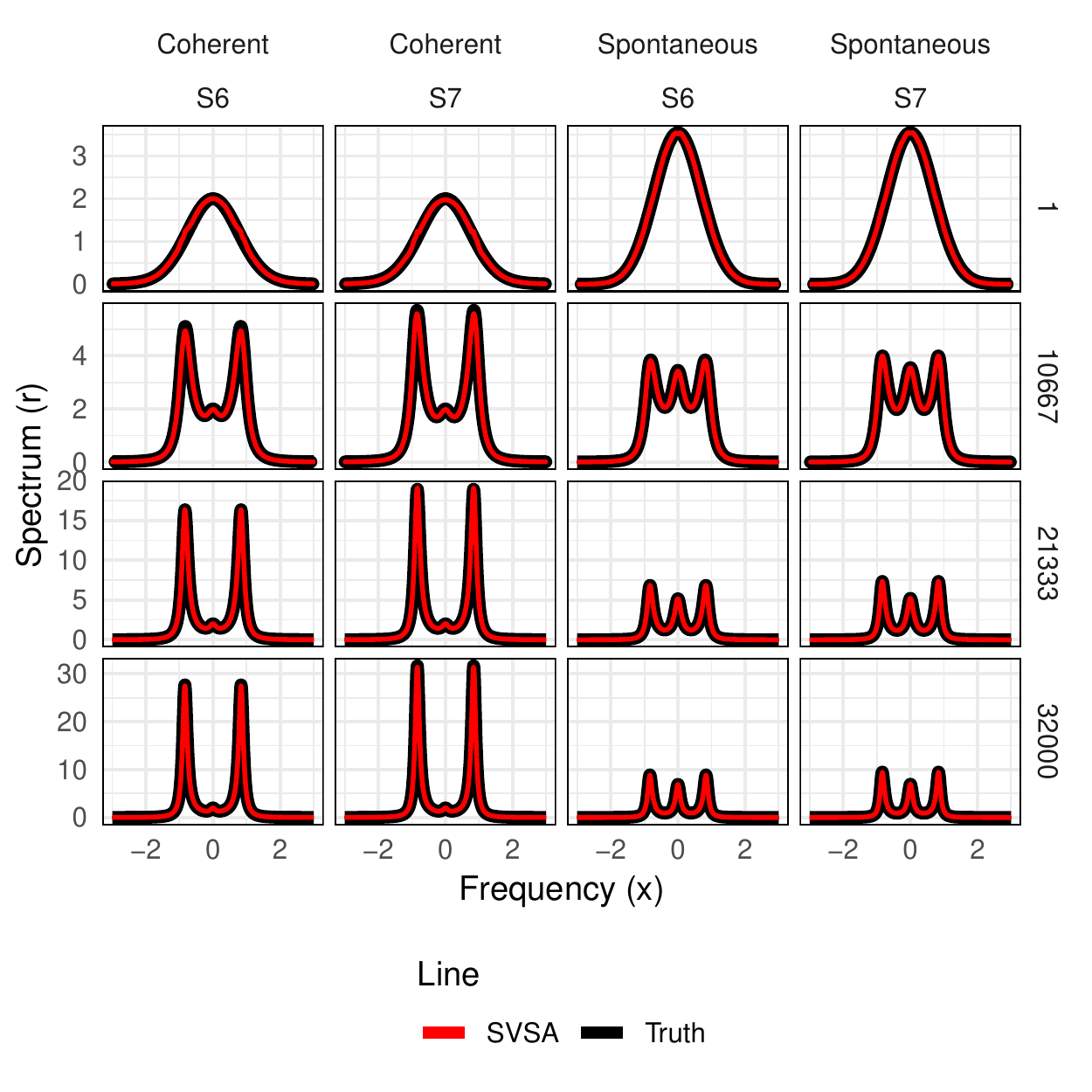}
  \caption{Examples of actual and predicted lineshapes for RBS. Black line is modeled line by the S6 or S7 model. The red line is the SVSA approximation.}\label{fig:ex_preds}
\end{figure}

\begin{figure}\centering
  \includegraphics[width=\figw]{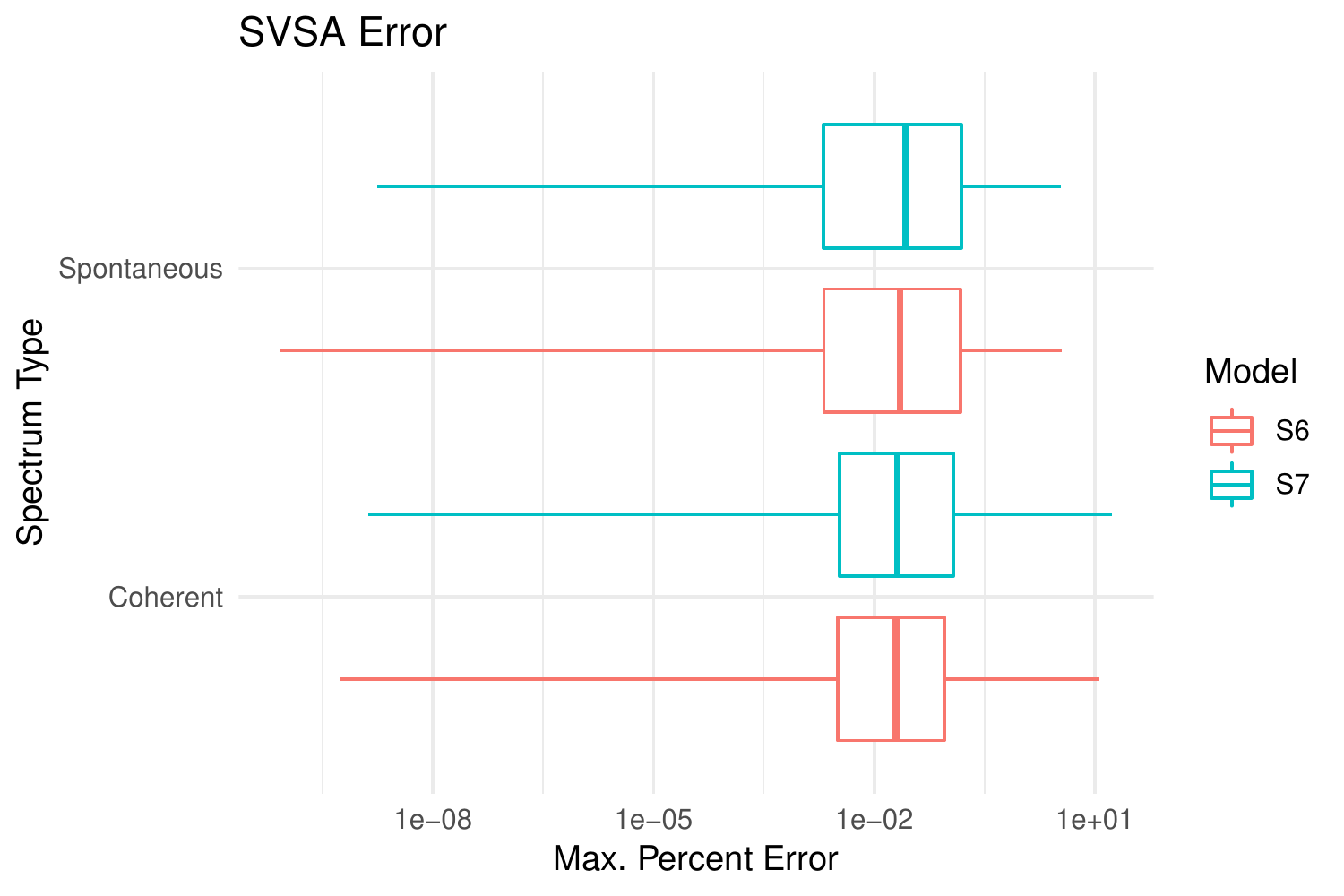}
  \caption{Percent Error for the four SVSA applications approximating the S6 and S7 models of coherent and spontaneous RB scattering.}\label{fig:boxplots}
\end{figure}

To test our method further, we build SVSA models to approximate coherent and spontaneous RB spectra using Tenti's S6 model and Pan's S7 model. We build these for models by training them on 500 example lineshapes over the same flow parameters previously used and setting $K=6$ and $\nu=0.1$. We then evaluate these approximators on 32,000 test lineshapes generated over the same parameter space. In Figure~\ref{fig:ex_preds}, we plot several example predicted lineshapes for our four models. The black lines are the true spectral lineshapes and the red lines are the estimated lineshapes by SVSA. In all cases, we can see that SVSA produces highly accurate approximations of the lineshapes. 
In Figure~\ref{fig:boxplots}, we plot boxplots of the percentage error for each of the four applications of SVSA across the 32,000 test examples. This figure shows that we have a median error of about $10^{-2}$ percent for the S6 and S7 models for both spontaneous and coherent scattering. These errors are well below the accuracy of the S6 and S7 models as compared to the true scattering lineshape. This demonstrates that SVSA is a versatile estimator that is applicable across models, like S6 and S7, and across scattering types, like coherent and spontaneous RBS. SVSA can approximate these different models with relatively little tuning and produce low error predictions across a large input parameter space. 

\section{Application}\label{sec:app}

To evaluate the efficacy of SVSA in a real application, we apply SVSA to a computational fluid dynamics (CFD) study of filtered Rayleigh scattering (FRS). FRS is an  experimental laser-diagnostic technique whereby the total energy $E$ of filtered Rayleigh-scattered light is used to quantify important properties of a flow like temperature, pressure, velocity, number density, etc. The data we use are Reynolds-Averaged Navier-Stokes CFD simulations of the supersonic shock wave dominated flow field in the experiment mentioned previously \citep{Baurle}. The output from these CFD simulations contains the aforementioned flow properties on a rectangular 3D grid of approximately 33 million points. Given these flow properties and the experimental setup, we can calculate the value of $E$ at each point. Doing this allows us to study the relationship between $E$, the flow properties, and the experimental setup. A better understanding of these relationships will aid in experiment optimization, interpretation, and analysis of the measured energy $E$ in future real FRS experiments. 

\begin{figure}\centering
  \includegraphics[width=\figw]{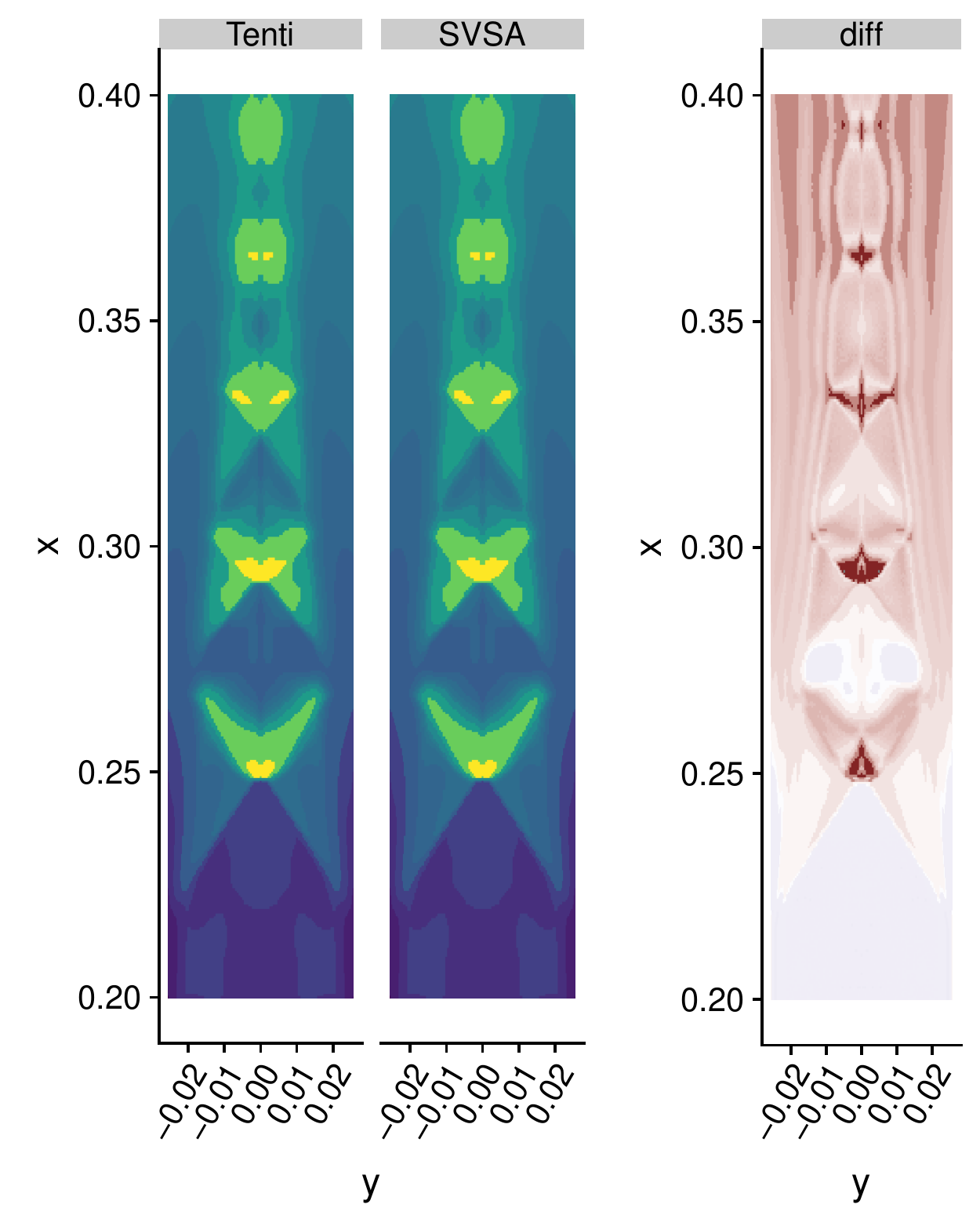}
  \includegraphics[width=\figw]{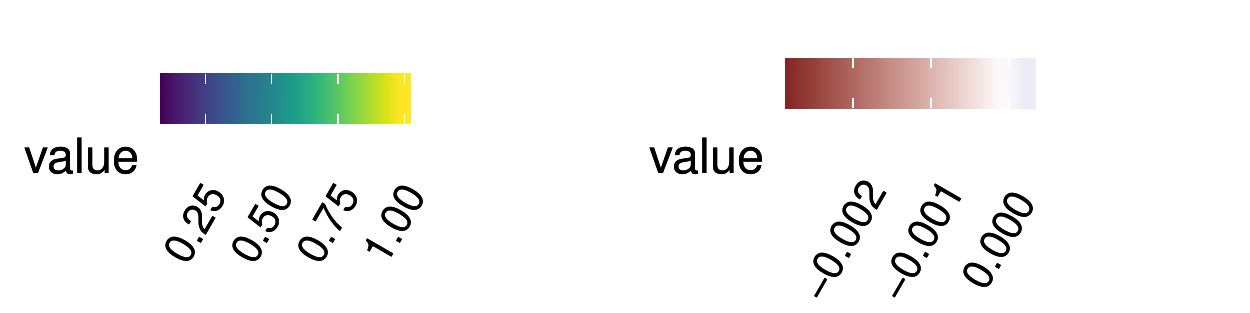}
  \caption{A single cross-sectional plane of $E$ calculated by Tenti and SVSA. The third plot is the difference between $E$ calculated using these two models. The energy has been normalized so that the maximum energy in the Tenti plot is one.}\label{fig:frs}
\end{figure}

To calculate $E$ at each of the 33 million points in our simulation, we need to know the flow parameters at each point (obtained from the CFD), the experimental setup (fixed in advance), and the RB scattering lineshape of the light at each point. To obtain the lineshape, one needs a lineshape model. Typically, Tenti's S6 model is used. However, because Tenti's model is comparatively slow, using Tenti's model to compute $E$ would take about 38 days to compute across all 33 million points. Alternatively, as our method is fast, using SVSA, it takes about $3\times 10^{-4}$ seconds to compute $E$ at each point and thus, less than three hours to compute across all 33 million points. In Figure~\ref{fig:frs}, we plot a single cross-sectional plane of $E$ calculated using the CFD data and either Tenti or SVSA. By eye, there is no discernible difference between Tenti and SVSA. Additionally, we plot the difference between the Tenti and SVSA calculated values of $E$. We can see from this plot that in absolute magnitude the difference between the two methods is never more than $2\times 10^{-3}$ (keeping in mind that the total scale for $E$ has been normalized to range between $0$ and $1$). Thus, SVSA produces a value of $E$ that is highly concurrent with Tenti's model but does so about 300 times faster. (While calculating the lineshape is 1-100 thousand times faster with SVSA, other parts of the calculation of $E$ are identical using Tenti and SVSA. Thus, the full calculation of $E$ doesn't see as large of an improvement as the calculation of the lineshape alone.)

\section{Conclusion}

In this paper, we developed a method called support vector spectrum approximation (SVSA) that uses machine learning to create efficient and accurate approximations of any existing spectral lineshape model in arbitrary flow regimes.The ability to produce approximations of existing lineshape models very efficiently is paramount for time-sensitive applications. In addition to our example of greatly speeding up a computational FRS study, we expect that SVSA will have application to real-time diagnostic techniques where efficiently calculating the lineshape from a continuous stream of data is necessary \citep{Yeaton2012,Binietoglou2016}. Additionally, we expect that SVSA will have application to problems outside of FRS and the RB spectra, for example, in approximating Raman spectra as part of CARS. We expect to explore such extensions in future work.

\section*{Funding}

National Aeronautics and Space Administration No.\\ NNX15AI20H subaward by the Virginia Space Grant Consortium No. 19-264-100527-010. 

\section*{Disclosures} The authors declare that there are no conflicts of interest related to this article.

\bibliographystyle{natbib}
\bibliography{frs,library}

\end{document}